\documentstyle[aps,eqsecnum,preprint,prd,epsf]{revtex}

\tighten
\begin{document}
\draft
\preprint{\parbox[t]{4.5cm}{TUM--T31--56/94/R\\hep--ph/9403407}}

\title{THE MASS DEFINITION IN HQET\\
AND A NEW DETERMINATION OF V$_{\text{cb}}$}
\author{Patricia Ball and Ulrich Nierste}
\address{Physik--Department/T30, TU M\"{u}nchen, D--85747 Garching, Germany}
\date{April 22, 1994}
\maketitle
\begin{abstract}
Positive powers of the mass parameter in a physical quantity 
calculated with the help of heavy quark effective theory  originate  
from a Wilson coefficient in the matching of QCD and HQET Green
function. We show that this mass parameter enters the calculation as 
a well--defined running current mass. 
We further argue that the recently found ill--definition of the pole
mass, which is the natural expansion parameter of HQET, does not 
affect a phenomenological analysis which uses truncated perturbative
series.
We reanalyse inclusive semileptonic decays of heavy mesons
and obtain the $c$ quark mass
$m_c^{\overline{\text{MS}}}(m_c) = (1.35\pm 0.20)\,\text{GeV}$ where
the error is almost entirely due to scale--uncertainties. We also obtain
$m_b^{\overline{\text{MS}}}(m_b) = (4.6\pm 0.3)\,\text{GeV}$ and
$|V_{cb}|(\tau_B/1.49\,\text{ps})^{1/2} = 0.036\pm 0.005$ 
where the errors come from the uncertainty in the kinetic energy
of the heavy quark inside the meson, in the experimental branching 
ratios, in QCD input parameters, and scale--uncertainties.
\end{abstract}
\clearpage

\section{Introduction}

During recent years the study of HQET, the effective theory of QCD
expanded in inverse powers of the heavy quark mass, has considerably
enlarged our understanding of low--energy  QCD (cf.\ \cite{reviews}
for reviews). Being successfully applied to a number of exclusive
decay processes of heavy--light systems containing one heavy quark and
one or two light quarks, it likewise supplies us with a number of
relations between static properties of such systems like the particle
spectrum and leptonic decay constants. Nevertheless, the question of
the correct definition of the heavy quark mass $m_Q$, the expansion
parameter of HQET, has played only a minor
r\^{o}le in the tremendous number of publications dealing with that theory.
When Green functions in full QCD are matched to 
those in HQET, one has to decide whether to identify $m_Q$ with the
{\em pole mass}\/ or with the renormalization scheme--dependent {\em current
mass}\/ of the QCD Lagrangian. 
Admittedly the question of which mass to use seems a rather academical
one from the viewpoint of phenomenology as long as the mass enters
explicitly only in inverse powers and implicitly in the strong 
coupling constant in the matching coefficients.
Yet the field of applications of HQET has expanded again and now
likewise encloses semileptonic inclusive decays of heavy mesons 
\cite{allg,dvf,rvf}. Two important
statements about inclusive decays could be obtained: first, in leading
order they are essentially free quark decays, and second, the leading
corrections to the free quark decay are of order $1/m_Q^2$. 
These results stimulated new determinations of the quark masses $m_c$
and $m_b$ and the CKM matrix element $|V_{cb}|$ from the experimental
measurements of the semileptonic branching ratios $B(D\to Xe\nu)$
and $B(B\to X_c e \nu)$, cf.\ 
Refs.~\cite{phenotypen1,phenotypen2,phenotypen3}. Since in
contrast to previous applications of HQET, the
inclusive decay rate depends on the heavy quark mass in its fifth$\,$(!)
power, a careful analysis of the correct definition of the mass
parameter is mandatory. In fact any {\em positive} power of the heavy 
mass originates from the matching of a QCD Green function to a HQET
Green function, so it is naturally a running current mass which enters
in this step. On the other hand the expansion parameter of the 
HQET Green function equals the pole mass \cite{grins,M2}, whose
precise  definition has recently been investigated by Braun and Beneke
and by Bigi et al.\ \cite{privet}. It was found that due to 
renormalon effects any attempt to define the pole mass beyond a 
finite order in perturbation theory is plagued by an 
intrinsic uncertainty of order $\Lambda_{\text{QCD}}$. 
 
The present paper is organized  as follows: 
In Sec.~II we  analyse the r\^{o}le of the mass parameter in the Wilson 
coefficient obtained from the matching of QCD and  HQET Green functions.
As a result we recover in the  leading order of the $1/m_Q$--expansion
the long--known renormalization scheme-- and scale--dependence of the parton 
model. We further argue that the ambiguity in the definition of the pole
mass, which is the expansion parameter of HQET, does not affect 
a phenomenological analysis which uses  the expansions in 
$1/m_Q$ and $\alpha_s$ only to a finite order.
In Sec.~III we investigate the inclusive decay rates of D-- and
B--mesons using the 
$\overline{\text{MS}}$ scheme.  
We  determine the quark masses and $|V_{cb}|$ in a manner that is very
similar to what was done in 
Refs.~\cite{phenotypen1,phenotypen2,phenotypen3}
and get a value of $|V_{cb}|$ that
is by about 10\% smaller than the results of
Refs.~\cite{phenotypen1,phenotypen2,phenotypen3}. 
This difference results solely from the transition of the 
on--shell scheme to the $\overline{\text{MS}}$ scheme. Special
attention is paid to the correct determination of the renormalization
scale--dependence. While it is huge in the inclusive rates itself, it
mainly drops out in $|V_{cb}|$.
Finally, in Sec.~IV, we discuss the results and draw some conclusions.

\section{The mass parameter in HQET} 

HQET  allows a  systematic expansion of QCD observables in inverse 
powers of a heavy quark mass. It   is  founded on the fact that 
the long--distance strong interaction of 
a heavy quark with mass
$m_Q$ is independent of its spin and flavour in the limit 
$m_Q \rightarrow \infty $. 
This heavy quark symmetry is violated by the finiteness 
of the heavy quark mass and by short--distance QCD interactions. 
The first correction is taken into account by including terms 
proportional to  powers of $1/m_Q$, while the effect of hard
gluons can be incorporated in a perturbation series in 
$\alpha _s (m_Q)$.
Usually these steps are treated independently: 
At first the heavy antiparticle field is integrated out in the 
path integral and the resulting expression is expanded in 
$1/m_Q$ yielding a series of local terms in the Lagrangian
\cite{mr}. 
Then Green functions in QCD and HQET are matched at a scale 
$\mu \approx m_Q$ 
at which both perturbative QCD and HQET are valid. 
In this step short--distance corrections are taken into account by
calculating loop corrections to the matching coefficients.  
Here one has to worry about the definition of the mass $m_Q$. 
The semileptonic inclusive decay rate discussed in Sec.~III 
contains $m_Q$ in the fifth power and the proper definition 
of $m_Q$ is of significant phenomenological importance. 

A mass parameter in a physical observable calculated with the help of HQET 
can  originate from two very different sources:
Consider  the matching of a QCD Green function 
$G_{\text{QCD}}$ to its HQET counterpart 
$G_{\text{HQET}}$ at some scale $\mu \approx m_Q$:
\begin{eqnarray}
G_{\text{QCD}} (m_Q,\mu) &=& C(m_Q,\mu) \left[  G^{(0)}_{\text{HQET}} ( \mu) 
              + \frac{1}{m_Q}   G^{(1)}_{\text{HQET}} (\mu) 
              + O \left( \frac{1}{m_Q^2}   \right) \right]
        \label{matchmu}.  
\end{eqnarray}
In Eq.\ (\ref{matchmu}) $m_Q$ can enter the Wilson coefficient $C$, which 
comprises the short--distance interactions
originating from scales 
larger than the matching scale $\mu$. 
The inverse masses  multiplying $G^{(k)}_{\text{HQET}}$, however, 
is the expansion 
parameter of HQET multiplying  the different orders of  
interaction operators in the HQET Lagrangian.   

The mass 
$m_Q$ in the Wilson coefficient $C$, 
which we will discuss first, 
is a short--distance quantiy and therefore 
clearly equals the {\em running current mass} $m_Q (\mu)$ in the 
renormalization scheme chosen for the calculation of 
$G_{QCD}$.
This fact becomes very transparent in the example  of a QCD Green function
whose Feynman graphs contain external heavy quark lines and internal 
lines corresponding to particles with masses 
$m_1 \gg m_2 \gg \ldots m_Q$:    
We first set  $\mu=\mu_1 \approx m_1$ and 
calculate the diagrams in the full standard model 
and in an effective field theory
 in which the heaviest internal particle is  
integrated out to  
  obtain some Wilson coefficient $C(m_1, \mu_1 )$.
Whenever we cross a particle threshold in the renormalization group 
evolution of $C$ to lower energies  
we have to repeat this factorization, which successively puts the masses 
of all internal heavy particles into the Wilson coefficient $C$. 
They  enter $C$ as running masses $m_i (\mu_i)$ evaluated at scales 
$\mu _i \approx m_i$. When we approach $\mu \approx m_Q$ 
we match the Green function 
in an effective QCD theory 
to a HQET Green function as displayed 
in (\ref{matchmu}), which adds $m_Q$ to the set of heavy masses 
appearing as arguments of $C$. There is no point in treating this mass 
differently from the others in $C$: The short--distance coefficient 
should not contain any information on  the long--distance interaction, 
i.e.\ whether the latter is encoded in a Green function 
calculated in HQET or in 
some other effective theory of QCD.

In a physical observable 
calculated with the help of HQET  {\em positive} powers of $m_Q$
obviously stem from the Wilson coefficient. 
In spite of this the inclusive decay rate discussed in Sec.~III
is usually entirely expressed in terms of the pole mass. 
The leading term in the $1/m_Q$--expansion is known to coincide 
with the parton model result, which has been obtained in  
\cite{cab} from an older QED calculation. 
In the latter an on--shell renormalization was used for the external fermion
so that the result was expressed in terms of the pole mass. 
In QCD this scheme is not  very adequate, because quarks are confined 
and their mass is not directly related to any observable.  
Nevertheless, as long as one stays within the parton model, 
which does not distinguish between the heavy meson and the 
heavy quark, 
the on--shell scheme is as good as any other scheme, to which one can 
easily pass by expressing the pole mass in the result by the
corresponding current mass.
By using HQET, however, we want to estimate corrections to the 
parton model result stemming from the binding of 
the heavy quark in the meson, and the correct tool to use is Wilson's 
operator product expansion (\ref{matchmu}). 

In the case of the inclusive decay rate the 
standard model   diagrams to be calculated for the 
left hand side of (\ref{matchmu}) are   
the heavy quark self--energy graphs
with the light quark and the lepton pair in the intermediate state. 
Its imaginary part determines the desired rate in the parton model.  
In general one would like 
 to include renormalization group improvement between 
the scales $\mu \approx m_W$ and $\mu \approx m_Q$ and would 
match the standard model diagram  first at $\mu \approx m_W$ 
to a corresponding diagram in an effective theory in which the 
W--boson is integrated out, which is then matched at 
$\mu \approx m_Q$ to the self--energy graph in HQET.  
In the case of the semileptonic decay rate, 
 (\ref{eq:Gamma}), however, no such improvement is
necessary due to the vanishing anomalous dimensions 
of the corresponding operators,
and the standard model diagrams are directly matched 
at $\mu \approx m_Q$ to the HQET Green function 
$\langle {\cal M}  | \overline{h} \Gamma h  | {\cal M}  \rangle$ 
and  to  matrix elements 
of the HQET operators subleading in $1/m_Q$.
Here ${\cal M}$  denotes the  heavy meson,  $h$ is the heavy quark
field in the effective theory and $\Gamma$ is the appropriate
Dirac structure.
In this step one gets the running mass of the heavy quark raised to the fifth 
power evaluated at the scale $\mu$. 
We stress that one is not forced to choose $\mu = m_Q$ exactly 
(see e.g.\ Neubert in \cite{reviews}). 
The choice of $\mu$ determines 
the separation of short and long--distance physics, the interaction 
from scales larger than $\mu$ is contained in the Wilson coefficient,
which contains the running quark mass.
 In an calculation to all orders
the result does not change with the variation of $\mu$. 
In practice one works with a truncated series and the dependence  
of the theoretical prediction 
on the scale $\mu $ estimates the 
reliability of the calculation. 
We will use this
tool extensively in the phenomenolocical analysis of Sec.~III.
The use of the pole mass in the Wilson coefficient
corresponds to the matching at an unnaturally low scale 
$\mu \approx 0.5\, m_Q$. In the case of the c quark this scale is too 
small for perturbative QCD to be trusted.

We will now discuss the nature of the mass parameter, whose inverse
powers  appears in the square brackets in (\ref{matchmu}) and stems 
from  the HQET Lagrangian.  

The starting point in the derivation of HQET is the decomposition of
the heavy quark's momentum $P_Q^\mu$ according to 
\begin{eqnarray}
P_Q^\mu &=& m_Q v^\mu + k^\mu, \label{hm}  
\end{eqnarray}
where $m_Q$ is the heavy quark mass, $v^\mu$ is the four--velocity of
the hadron and $k^\mu$  is the residual momentum. 
The latter is usually constrained by the condition
{\samepage
\begin{eqnarray}
k^\mu = {\cal O}(\Lambda_{\text{QCD}}) & \simeq &
\text{const.,}   \label{old}      \\[-18pt]
& \scriptstyle m_Q \to \infty &  \nonumber
\end{eqnarray} }
so that finally physical observables $X$ are expanded in a 
power series in 
$\Lambda_{\text{QCD}}/m_Q$:
\begin{eqnarray}
X & \sim & \sum_k a_k 
    \left( \frac{\Lambda_{\text{QCD}} }{m_Q} \right)^{k} . \label{hqe} 
\end{eqnarray}

In Eq.\ (\ref{old}) we have emphasized that in HQET the constraint   
$k^\mu={\cal O}(\Lambda_{\text{QCD}})$ is understood as a scaling relation: 
The residual momentum does not diverge when the limit 
$m_Q \rightarrow \infty $ is performed.  
Obviously (\ref{old}) cannot hold for every definition of the mass,
because a change in the
QCD renormalization prescription for $m_Q$ 
may 
redefine $k^\mu$
in (\ref{hm}) by an amount 
involving $\alpha _s (m_Q) m_Q v^\mu $, which scales like 
$m_Q / \ln m_Q $ rather than staying constant. 

Authors addressing  the definition of the mass 
indeed only allow  on--shell--like  renormalization conditions for 
$m_Q$ \cite{grins,M2,uab}. 
Let us recall their key arguments:
The QCD quark self--energy 
\begin{eqnarray}
\Sigma &=& 
m_Q  \Sigma_1 +( p\hspace{-5.5pt}/\hspace{5.5pt}\!\! -m_Q) \Sigma_2 
\end{eqnarray}
can be sandwiched between two projectors 
$P_v^+=(1+v\hspace{-5.5pt}/\hspace{5.5pt}\!\!)/2$ 
and be expanded as 
\begin{eqnarray}
\lefteqn{ P_v^+ \Sigma P_v^+ \; =  \; P_v^+ 
    \left[ m_Q \Sigma_1 + v \cdot k \, \Sigma_2  \right] P_v^+}  \nonumber\\
&=& \frac{\alpha _s C_F}{4 \pi}  
    \left[ 
  m_Q \Sigma_{10} +v \cdot k \left( \Sigma_{11} + \Sigma_{20} \right) 
   +  {\cal O} \left( \frac{ ( v \cdot k) ^2 }{m_Q} \right) 
     \right] P_v^+ 
   + {\cal O} \left( m_Q \alpha_s^2    \right) . \label{self} 
\end{eqnarray}
Here the one--loop contribution to 
$\Sigma_i$ is expanded 
with respect to $v \cdot k/ m_Q$ and 
$\alpha _s C_F /( 4 \pi) \Sigma_{ij}$ 
is the $j$-th coefficient.
Now 
$\Sigma$ has to match  
the HQET self--energy $v\cdot k \, \tilde{\Sigma}$ 
up to ${\cal O}( v \cdot k / m_Q)$.
For this  
an on--shell--like renormalization condition
\begin{eqnarray}
\Sigma_1|_{m_Q^2-P_Q^2 = \rho}&=& 0 \label{osl}
\end{eqnarray}
with $\rho \approx 0$ fixed as
$m_Q \rightarrow \infty$ was required in \cite{grins}. 
Otherwise the term $m_Q \Sigma_{10}$ in (\ref{self}) would diverge 
as $m_Q \rightarrow \infty $ and thereby could not be matched to any  
HQET Green function, which scales at most as a constant due to (\ref{old}). 
In the following we set $\rho =0$, so that
$m_Q$ in (\ref{hm}) equals the pole mass, when the constraint 
(\ref{old}) is imposed on $k^\mu$.

{}From a phenomenologist's point of view the appearance of a quark 
pole mass is unsatisfactory, because quarks do not exist 
as free particles, so that the pole mass  is no observable. 
Any QCD calculation  relates observables to running current masses,
which are contained in  Wilson coefficients. 
The pole mass has to be  {\em calculated} in terms of  the current mass
extracted from  some experiment. 
Eq.\ (\ref{hm}) implies that the pole mass is calculated to all
orders in perturbation theory: 
By  truncating the perturbative series
defining $m_{\text{pole}}$ at order $n$ one picks up an error of the
order $\alpha_s^{n+1} (m_Q) m_Q = {\cal O}(m_Q/ \ln^{n+1} m_Q)$ in the residual
momentum, so that it would not 
stay  constant in the limit $m_Q \rightarrow \infty$.  
Moreover, the authors of \cite{privet} have proven that the pole mass 
suffers from an extra infrared renormalon
situated at $u=1/2$ in the Borel plane, so that the result depends 
on the summation prescription chosen for the divergent perturbative 
series and  
is ill--defined by a term of  order $\Lambda_{\text{QCD}}$.  

We therefore want to avoid the use of the pole mass in the first step 
of the construction of HQET and use an arbitrary current mass instead.
The final result, however, will be the same as with the pole mass as 
the starting point, but an alternative  derivation might be illustrative. 

If $m_Q$ denotes the current mass in an arbitrary renormalization
scheme of QCD, (\ref{old}) can no more be imposed on $k^\mu$ and we
have instead 
{\samepage \begin{eqnarray} 
k^\mu & \simeq & {\cal O}(\alpha _s (m_Q) m_Q) 
    \, = \, {\cal O} \left( \frac{m_Q}{\ln (m_Q/\Lambda_{\text{QCD}})}
\right), \label{new}        \\[-20pt]
& \scriptstyle m_Q \to \infty & \nonumber
\end{eqnarray}}   
because a change of the renormalization prescription for $m_Q$
redefines $k^\mu$ by a quantity involving 
$\alpha _s (m_Q) m_Q = {\cal O}(m_Q / \ln m_Q)$.  
Nevertheless 
the heavy quark expansion of QCD Green functions remains  reasonable,
because one still has $k^\mu / m_Q \to 0$ for \mbox{$m_Q \to \infty$},
so that $v \cdot k/m_Q$ is a small parameter.
The heavy quark field $h_v (x) = P_v^+ Q(x) \exp (i m_Q v\cdot x) $
is now unambiguously defined, because it involves the current mass.

We can now investigate how the matching of QCD and HQET has to be
performed in our  approach: 
Consider the (non--truncated) two--point function $G_{2}$ in QCD. 
The tree--level part reads
\begin{eqnarray}
P_v^+ G_2^{(0)} P_v^+ &=& i P_v^+ \frac{p\hspace{-5.5pt}/\hspace{5.5pt}\!\!+
m_Q}{p^2 -m_Q^2} P_v^+ \nonumber\\
& = & \frac{i P_v^+ }{v \cdot k} + \frac{i P_v^+}{2 m_Q} 
              \left[ 1- \frac{k^2}{(v \cdot k)^2 }   \right]
            +{\cal O} \left( \frac{1}{m_Q \ln m_Q }  \right). \label{ld}
\end{eqnarray}  
In the ${\cal O}(\ldots )$--part 
we have taken into account 
that $k^\mu$ scales according to
(\ref{new}). 
The ${\cal O} \left( \alpha _s  \right)$--part yields for $\mu=m_Q$:
\begin{eqnarray}
\lefteqn{
P_v^+ G_2^{(1)} P_v^+ \; = \;  - i P_v^+ \frac{p\hspace{-5.5pt}/\hspace{5.5pt}\!\!+
m_Q}{p^2 -m_Q^2} 
                     \Sigma  \frac{p\hspace{-5.5pt}/\hspace{5.5pt}\!\!+m_Q}{p^2 -
m_Q^2} P_v^+ } \nonumber\\ 
&=& -i P_v^+ \frac{\alpha _s (m_Q) C_F}{4 \pi}  \left[ \frac{m_Q}{(v \cdot k)^2}
                               \left( \Sigma_{10} -
                              \delta_m \right) \right. \nonumber\\  
&& \left.      + \frac{1}{v \cdot k} 
       \left( \Sigma_{11} + \Sigma_{20} +\delta_2 + 
              \left( 1- \frac{k^2}{(v\cdot k)^2} \right) 
               \left( \Sigma_{10}-\delta_m   \right)  \right)
                \right] 
           + {\cal O} \left( \frac{1}{m_Q \ln m_Q} \right) , \label {sld}
\end{eqnarray}  
where $-\delta_m C_F \alpha _s /( 4 \pi)$ and 
 $\delta_2 C_F \alpha _s /( 4 \pi)$ are the mass and wave function
counterterms in the QCD  Lagrangian. 
Eqs.\ (\ref{ld}) and (\ref{sld}) illustrate that with (\ref{new}) the 
$1/m_Q$--expansion now yields a series in which each term is still
suppressed  compared to the preceding one, but only by a factor 
of order $1/\ln ( m_Q / \Lambda_{\text{QCD}})$ rather than of order 
$\Lambda_{\text{QCD}}/m_Q$. 

The new feature in the matching 
of (\ref{ld}) and (\ref{sld}) to the corresponding expressions in
HQET is the appearance of the term $\Sigma_{10}-\delta_m$ in
(\ref{sld}) which equals zero in the on--shell scheme.
To accomodate for it  we need an additional term 
\begin{eqnarray}
-\Delta m \bar{h}_v h_v \nonumber 
\end{eqnarray}
in the HQET Lagrangian with
\begin{eqnarray} 
\Delta m &=& \frac{\alpha _s C_F}{4 \pi} m_Q 
             \left( \delta_m -\Sigma_{10} \right) 
             + {\cal O} \left( m_Q \alpha _s^2 \right).  \label{et}     
\end{eqnarray}
The HQET two--point Green function 
with one insertion of $-\Delta m \bar{h}_v h_v $ then matches the
first term in (\ref{sld}).
When working to $n$-th order in $\alpha _s$,
(\ref{et}) must be adjusted to cancel the $m_Q \Sigma_1$--terms
to order $\alpha _s^n$.
Clearly  $\Delta m$ in the above one--loop example, Eq.\ (\ref{sld}) ,
is nothing but $m_{\text{pole}}^{(1)}-m_Q$. In the 
$\overline{\text{MS}}$ scheme
\begin{equation}\label{eq:deltamMS}
\Delta m (\mu) = \frac{\alpha_s (\mu) }{\pi} m_Q 
       \left( \frac{4}{3} + \ln \frac{\mu^2 }{m_Q^2} \right) . 
\end{equation} 
$\Delta m $ is a purely short--distance quantity and does not require 
any definition of $m_{\text{pole}}$ beyond perturbation theory. Consequently 
we treat the ``residual mass term'' $-\Delta m \bar{h}_v h_v$ 
as an interaction vertex and not as a part
of the kinetic Lagrangian. 
The other term involving $\Sigma_{10}-\delta_m$ 
in (\ref{sld}) stems from the $1/m_Q$--part of the 
propagators between which $m_Q  \Sigma_1$ is sandwiched. They clearly 
match the HQET two--point function with one insertion of 
$-\Delta m \bar{h}_v h_v$ 
and one insertion of the sum of the usual $1/m_Q$--subleading operators  
$1/( 2 m_Q) \bar{h}_v (i D)^2  h_v$  and 
$ - 1/( 2 m_Q) \bar{h}_v (i v \cdot D )^2  h_v$.
It is evident 
how  the matching of $m_Q \Sigma_{10}-\delta_m$ 
to $-\Delta m $ 
works to higher orders in $v \cdot k/m_Q$.
The matching of the wave function counterterms is not different from 
the conventional approach with $\Delta m=0$.

The difference between the residual mass term introduced in 
\cite{res} and $\Delta m$ in (\ref{et}) is that the former was
constrained to be of order $\Lambda_{\text{QCD}}$ while 
$\Delta m $ is of order $\alpha_s m_Q$. The appearance of a 
positive power of $m_Q$  
in the HQET Lagrangian 
has a dramatic consequence for the 
$1/m_Q$-expansion of Green functions: 
With the use of the pole mass inverse powers of $m_Q$ and 
logarithms of $ m_Q$ 
are neatly separated, so that the $1/m_Q$--expansion coincides
term--by--term with 
an operator product expansion in $\Lambda_{\text{QCD}}/m_Q$ 
as in (\ref{hqe}). 
By using a current mass as the expansion parameter and introducing 
$\Delta m_Q$ in the HQET Lagrangian we have reshuffled 
the two expansions in $1/m_Q$ and $\alpha_s (m_Q)$. 
In order to arrive directly at an operator product expansion 
(\ref{hqe}) one  necessarily has to separate the heavy quark mass 
and self--interaction from the interaction with the 
light degrees of freedom, which is of order $\Lambda_{\text{QCD}}$.  
The authors of \cite{privet} have
shown that this separation is ambiguous beyond perturbation theory 
and can introduce a residual mass term in the HQET Lagrangian 
which is of order $\Lambda_{\text{QCD}}$  and is a remnant of 
the long--distance self--interaction of the heavy quark. As for the 
short--distance contributions to the self--energy  of the heavy quark, 
this separation is, however, possible and mandatory to achieve 
the abovementioned separation of powers and logarithms of 
$m_Q$ directly.\footnote{We thank M.\ Beneke for clarifying this point.}
 
By using an arbitrary current mass as the expansion parameter we  have
put  the  quark self--energy in the chosen QCD scheme entirely 
into the HQET Lagrangian.  

Nevertheless one can easily recover the correct operator product 
expansion: 
To this end we 
investigate how 
 the hadronic parameters of HQET transform under a change  of  
the renormalization scheme  used to define $m_Q$.

Consider two schemes differing by a finite 
renormalization of the mass: 
\begin{eqnarray}
m_Q ^\prime &=& z_m m_Q \;=\; m_Q+ \delta m_Q \nonumber\\ 
\Delta m_Q ^\prime &=& \Delta m_Q- \delta m_Q.    \label{schtrafo}
\end{eqnarray}
The choice $\Delta m_Q=\delta m_Q$ transforms to the pole mass.
The transformation (\ref{schtrafo}) modifies the HQET quark field as 
\begin{eqnarray}
h_v ^\prime (x) &=&  h_v e^{i v x \delta m_Q} .  \label{schtrafo2} 
\end{eqnarray}
The 
parameter $\bar{\Lambda}$ measures the  mass  
difference between meson and quark. 
One easily finds:
\begin{eqnarray}
\bar{\Lambda} ^\prime &=& \frac{i v \partial 
\langle 0| \overline{q^\prime} \Gamma h_v^\prime | M(v)  \rangle}{\langle 0| 
           \overline{q}^\prime \Gamma h_v ^\prime | M(v)  \rangle}, \nonumber\\
&=& \bar{\Lambda} - \delta m_Q , \label{example}
\end{eqnarray}
so that the meson mass 
\begin{eqnarray}
M_Q&=& m_Q+\bar{\Lambda}+{\cal O} \left( 
      \frac{\Lambda^2_{\text{QCD}}}{m_Q} \right) \label{mesmas}
\end{eqnarray}
is scheme--independent. 
In an arbitrary scheme $\bar{\Lambda}$ contains the quark self--energy,
whose short--distance part depends on $m_Q$, so that we 
have to   add a flavor label to 
 $\bar\Lambda $ and write in the e.g.\ 
$\overline{\text{MS}}$ scheme for $\mu=m_Q$: 
\begin{eqnarray}
\bar{\Lambda}_Q^{\overline{\text{MS}}} &=& 
     \bar{\Lambda}_{\text{pole}}  
      + \frac{4}{3} \frac{\alpha_s (m_Q) }{ \pi} m_Q 
         + {\cal O} \left(\alpha_s^2 m_Q \right)  , \label{laq}  
\end{eqnarray}
where $\bar{\Lambda}_{\text{pole}}$ is the usual
$\bar{\Lambda}$--parameter which corresponds to the choice of 
$m_Q=m_{\text{pole}}$ 
and obeys the heavy quark symmetry. 
So far we have just trivially pushed self--energy contributions from 
the mass into $\bar{\Lambda}_Q$. Reference to the pole mass is for
the first time made in (\ref{laq}), which expresses that only with 
an on--shell renormalization for $m_Q$ the quantity 
$\bar{\Lambda}$ is of order $\Lambda_{\text{QCD}}$. 
{}From \cite{privet}  we know that the separation of 
$m_{\text{pole}}$ and $\bar{\Lambda}_{\text{pole}}$ 
is ambiguous by long--distance terms of order $\Lambda_{\text{QCD}}$.
With our formalism we clearly do not resolve this ambiguity, but just 
make clear that it does not affect a practical phenomenological
analysis, where one would like to use (\ref{mesmas})
to extract, say, $\bar{\Lambda}_{\text{pole}}$ 
from an analysis of the D--meson 
system and to insert it into a prediction for the B--meson system. 
In the first step one extracts the current charm quark mass from 
some observable which is calculated with radiative corrections to 
some order $\alpha_s^n$. Clearly in (\ref{laq})  one will only subtract 
the quark self--energy to the same order, so that  the prediction   
for the calculated observable will then have an error of the order 
$\alpha_s ^{n+1} $. 
One is never faced with the problem to define the pole mass beyond 
perturbation theory.
Even if we did the calculation 
to a very high order, so that we saw the 
renormalon--induced divergence of the
perturbative series 
for the self--energy in Eq.\ (\ref{laq}), we would naively expect that the 
renormalon ambiguity cancels in the prediction, if we fixed the
summation prescription for the perturbation series in both the 
D- and B- analysis in the same way. 

We will next discuss the scheme--dependence of the parameters 
$\lambda_1$ and $\lambda_2$, which 
appear at order $1/m_Q$ and 
parametrize the matrix elements with two heavy meson states of equal
velocity.
At this order the residual mass $\Delta m$ has to be taken into
account. We first want to define $\lambda_1$ and $\lambda_2$ in a
scheme--invariant way, i.e.\ such that they are independent 
of $\Delta m$.
{}From \cite{res} we know that the latter can be obtained
by replacing the covariant derivative $D^\mu$ by $D^\mu+i v^\mu \Delta
m$.
Indeed, for a heavy meson ${\cal M}$ with mass $M_Q$  we find 
\begin{eqnarray}
2 M_Q \lambda_1 & = & \langle\,{\cal M} \,|\,\bar{h}_v (iD-v \Delta m)^2
 h_v\, | \,{\cal M} \,\rangle (1+{\cal O}(1/m_Q)), \label{lambada} \\
6 M_Q \lambda_2 (\mu ) & = & \langle \,{\cal M} \,|\,\bar{h}_v \frac{g}{2}\,
\sigma_{\mu\nu} F^{\mu\nu} h_v\,|\, {\cal M} \,\rangle(1+{\cal O}(1/m_Q))
\nonumber
\end{eqnarray}
to be invariant with respect to the transformations (\ref{schtrafo}) and 
(\ref{schtrafo2}). Consider now some observable whose $1/m_Q$--part looks for 
$m_Q=m_{\text{pole}}$ like 
\begin{eqnarray}
\frac{\lambda_1}{m_{\text{pole}}} 
      \left( a+ b \alpha_s + {\cal O}\left( \alpha^2_s \right)\right) \label{1m} 
\end{eqnarray} 
with some constants $a$ and $b$ plus a similar term with $\lambda_2$. 
In an arbitrary scheme one gets the current mass instead in the
denominator of (\ref{1m}), but the second  order in the 
$1/m_Q$--expansion involves matrix elements with 
one insertion of  $\Delta m$ and  
two insertions of
the $1/m_Q$--operators. 
Since 
$\Delta m/m_Q= {\cal O}(\alpha_s )$, these contributions are of the same order
as the first order radiative corrections in (\ref{1m}). 
Renormalization scheme--invariance of QCD requires that both terms 
combine to a scheme--invariant quantity:
\begin{eqnarray}
\frac{1}{m_Q} - \frac{\Delta m }{m_Q^2 } &= &\frac{1}{m^{(1)}_{\text{pole}}} 
 \left( 1+ {\cal O} ( \alpha^2_s ) \right).  
\end{eqnarray} 
If we want to include explicit radiative corrections to order $n$ in 
(\ref{1m}), we must calculate $\Delta m$ to order $\alpha^n_s$ and 
have to take into account matrix elements with multiple  insertions of 
$\Delta m$ up to order $1/m_Q^{n+1}$, so that  the reordering of the 
expansions in $1/m_Q$ and $\alpha_s$ takes place over $n$ orders 
in the $1/m_Q$--expansion.   
Hence the final result is the same as with the use of the pole mass
from the very beginning. Yet in our derivation  
the pole mass appears explicitly only 
in the last step of the derivation. 
One has to calculate it only to the same (finite) order in 
perturbation theory as the explicit radiative corrections to the
corresponding term in the observable (\ref{1m}) under consideration. 

In the sketch given above the ambiguously defined pole mass to all orders 
does not appear in the definition of the 
HQET field $h_v$, instead the self--energy $-\Delta m$ 
of the quark is encoded order by order in $\alpha_s$ in the 
HQET Lagrangian.
We have made plausible that the use of HQET in a phenomenological 
analysis, which includes radiative corrections to some  {\em finite} order 
to the various terms in the $1/m_Q$--expansion, does not require 
the calculation of $m_{\text{pole}}$ beyond the same order in 
its perturbative series.  

Let us emphasize again that our formalism of HQET with the current
mass as the starting point leads exactly to the same result 
as the conventional one, if one uses the pole mass calculated
perturbatively to the same order as the explicit radiative
corrections multiplying the $1/m_Q$--term  under consideration.
Our statements concerning the r\^{o}le of the mass appearing in the 
Wilson coefficients as, e.g.,\ the $m_Q^5$--term in the inclusive decay
rate, lead, however, to numerical effects in phenomenological analyses
as demonstrated in Sec.~III.

\section{Semileptonic inclusive decays}

Let us now turn to an important application of the considerations of
the last section. Quite recently, the heavy quark expansion has been
extensively applied to the problem of inclusive heavy meson decays,
both semi-- \cite{allg,dvf,rvf} and nonleptonic ones \cite{nonlep}. 
Two major results were obtained, namely first that inclusive decays 
are determined by the free quark decay in leading order in the heavy 
quark expansion, and second that non--perturbative corrections to the 
free quark decay picture are suppressed by terms of order $1/m_Q^2$; 
there are no terms of order $1/m_Q$. These results allow an immediate
application to the determination of the CKM matrix element $|V_{cb}|$.
This task was tackled in the papers 
Ref.\ \cite{phenotypen1,phenotypen2,phenotypen3} where the following
procedure was adopted: first the $c$ quark mass $m_c$ is determined
from the branching fraction $B(D\to Xe\nu)$ in dependence on the
unknown non--perturbative parameters. Then HQET is invoked in order to
fix the $b$ quark mass $m_b$ from the known value of $m_c$. Insertion
in the theoretical expression for $B(B\to X_c e\nu)$ finally yields
$|V_{cb}|$ \cite{phenotypen1,phenotypen2}, or, assuming $|V_{cb}|$ as
known from other sources, bounds on the non--perturbative parameters
and quark masses, respectively, cf.\ 
\cite{phenotypen3}. In the present analysis we will closely follow the
sketched method, but work in the $\overline{\text{MS}}$ scheme and
keep track of possible scheme--dependences.

The non--perturbative corrections to the free quark decay picture 
can be expressed in terms of two matrix elements,
\begin{eqnarray}
2m_B\lambda_1 & = & \langle\,B\,|\,\bar{b}_v (iD-v\Delta m)^2
b_v\,|\,B\,\rangle (1+{\cal O}(1/m_b)), \nonumber\\
6m_B\lambda_2(\mu_b) & = & \langle\,B\,|\,\bar{b}_v \frac{g}{2}\,
\sigma_{\mu\nu} F^{\mu\nu}b_v\,|\,B\,\rangle(1+{\cal O}(1/m_b)),
\end{eqnarray}
where $b_v$ is defined as $b_v = e^{im_b v x}b(x)$, $b(x)$ being the 
$b$ quark field in full QCD, and $v_\mu$ is the four--velocity of
the B meson. The corrections of order $1/m_b$ on the right--hand 
side account for the fact that physical states in QCD and HQET differ
by terms of order $1/m_Q$. The above
quantities have an immediate physical interpretation by virtue of the 
scheme--invariant relation
\begin{equation}\label{eq:massen}
m_B = m_b + \bar\Lambda_b -\frac{\lambda_1+3\lambda_2(\mu_b)}{2m_b} + 
{\cal O}\left(\frac{\alpha_s}{m_b},\frac{1}{m_b^2}\right)
\end{equation}
where $\lambda_1/(2m_b)$ plays the r\^{o}le of the kinetic energy of the
heavy quark's Fermi motion inside the meson and the term in
$\lambda_2$, which has non--zero anomalous dimension in HQET,
accounts for its spin--energy in the chromomagnetic
field. $\mu_b$ is the matching scale of HQET onto QCD. 
$\bar\Lambda_b$ can be interpreted as the binding energy of the
light degrees of freedom in the meson. From (\ref{eq:massen}) one
readily infers
\begin{equation}
\lambda_2(\mu_b) \approx \frac{1}{4}(m_{B^*}^2-m_B^2)\approx 0.12\,
\text{GeV}^2
\end{equation}
which is true in any renormalization scheme.
Unfortunately, $\lambda_1$ is no observable, but has to be determined
within some model--calculation. Recently, the value $\lambda_1 =
-(0.6\pm 0.1)\,\text{GeV}^2$ was obtained from QCD sum rules \cite{BBlast}. 
In view of the criticisms raised in \cite{vir}, we nevertheless
prefer to leave $\lambda_1$ as an open parameter to be varied within
the interval $[-0.7,0]\,$GeV$^2$ where the upper bound is taken from
\cite{dvf}, the lower one from the sum rule determination \cite{BBlast}. 

\begin{figure}[h]
\centerline{
\epsfysize=0.24\textheight
\epsfbox{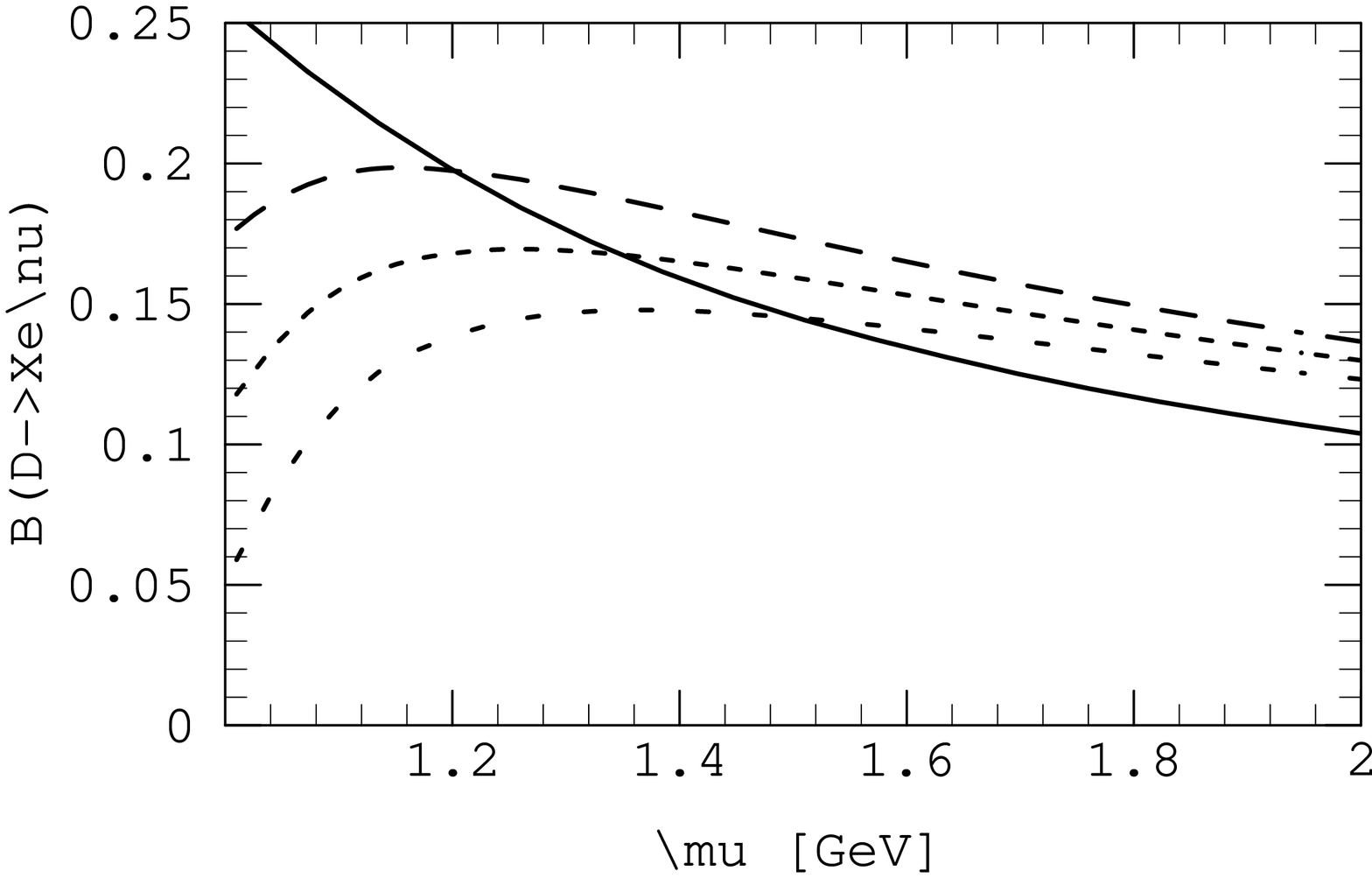}}
\vspace{0.1in}
\caption[]{The branching ratio $B(D\to Xe\nu)$ as function of the 
renormalization scale $\mu$ for $m_c(m_c)=1.35\, \text{GeV}$, 
$\lambda_1=0\,\text{GeV}^2$, $\lambda_2(m_c) = 0.1\,\text{GeV}^2$. 
Solid line: $B=\tau_D\Gamma$ calculated according to Eq.\ 
(\protect{\ref{eq:Gamma}}), dashed lines: 
$B$ according to Eq.\ (\protect{\ref{eq:Gamma}}) with inclusion of 
${\cal O}(\alpha_s^2(\mu))$ corrections according to Eq.\ 
(\protect{\ref{eq:alphaquadrat}}). The different dashed lines
correspond to three different values of $c_3$: $10$, $0$, $-10$ (from
above).}\label{fig:scalesGammac}
\vspace{0.1in}
\centerline{
\epsfysize=0.45\textheight
\epsfbox{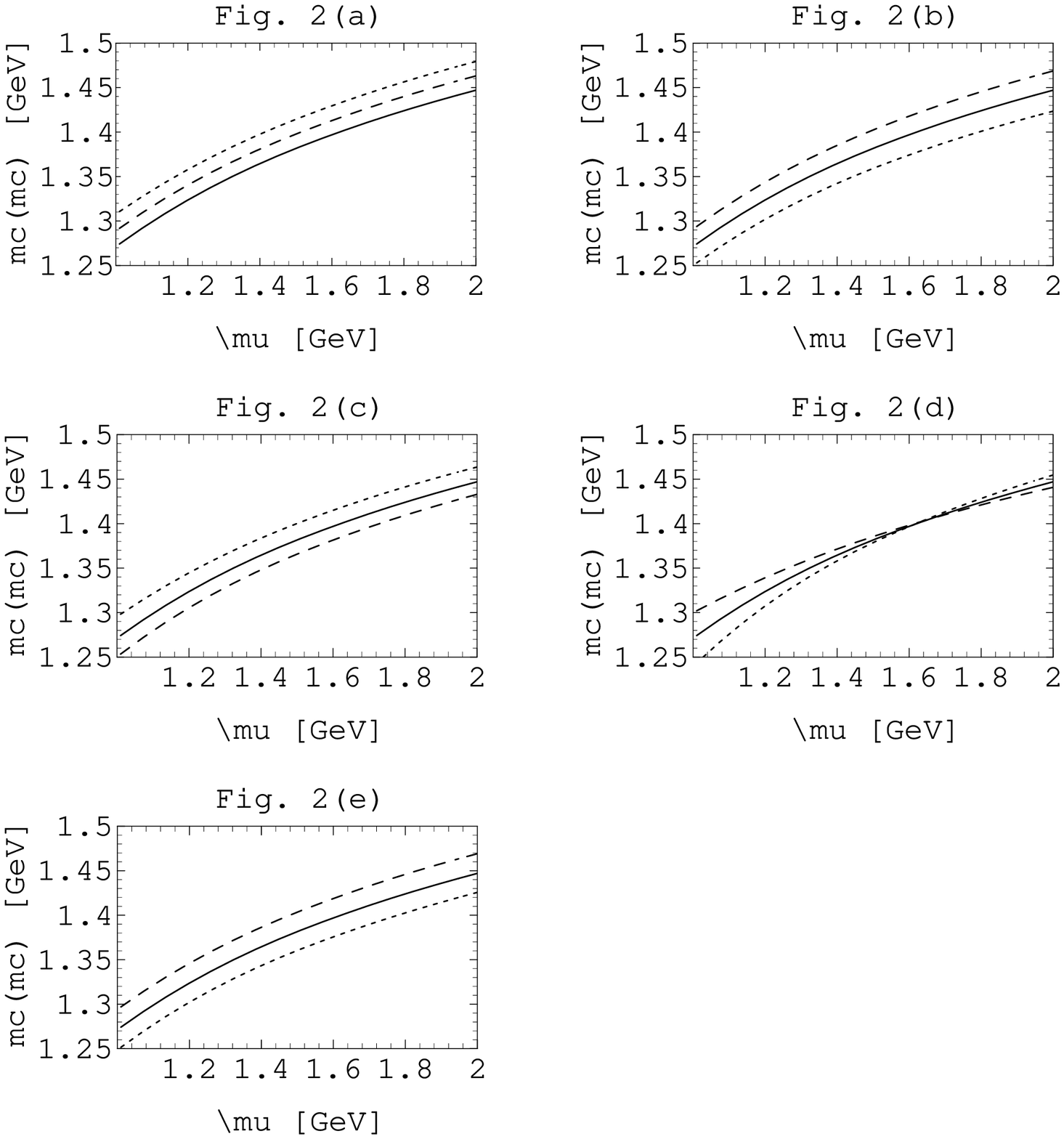}}
\vspace{0.1in}
\caption[]{$m_c(m_c)$ as function of the renormalization scale $\mu$ for
different values of the input parameters. The solid lines are
calculated with $\lambda_1 = 0\,\text{GeV}^2$, $B(D\to Xe\nu)= 0.172$,
$m_s(1\,\text{GeV}) = 0.2\,\text{GeV}$, 
$\Lambda^{(4)}_{\overline{\text{MS}}} = 300\,\text{MeV}$.
The (long-, short-) dashed lines are obtained by replacing the
respective parameter by (a) $\lambda_1 = (-0.35,-0.7)\,\text{GeV}^2$, 
(b) $B(D\to Xe\nu) = (0.191,0.153)$, (c) $m_s(1\,\text{GeV}) =
(0.15,0.25)\, \text{GeV}$, (d) $\Lambda^{(4)}_{\overline{
\text{MS}}} = (250,350)\,\text{MeV}$. In (e) we have
included $1/m_c^3$ corrections as described in the text.}\label{fig:mc}
\end{figure}

The heavy quark expansion of the decay rate of the semileptonic decay 
$B\to X e \nu$ with unobserved $X$ in the
final state\footnote{In the following we identify $B\to X e \nu$ with
$B\to X_c e \nu$ where the $b$ quark decays into a $c$ quark since 
charmless decays $B\to X_u e \nu$ can safely be
neglected because of $|V_{ub}/V_{cb}|^2\lesssim 1\% $ \cite{PD}.}
then reads \cite{allg,cab}:
\begin{eqnarray}
\Gamma(B\to X_c \ell \nu) & = & \frac{G_F^2 (m^R_b)^5}{192\pi^3}\,
|V_{cb}|^2 \, \left[ \left\{ 1-\frac{2}{3}\,\frac{\alpha_s^R}{\pi}\,
g^R\!\left(\frac{m_c^R}{m_b^R}\right) + \frac{\lambda_1}{2(m_b^R)^2}\right\}
f_1\!\left(\frac{m_c^R}{m_b^R}\right) \right.\nonumber\\
& & \phantom{\frac{G_F^2 m_b^5}{192\pi^3}\,
|V_{cb}|^2 \;\;\;} \left. {}- \frac{9\lambda_2}{2(m^R_b)^2}\,
f_2\!\left( \frac{m_c^R}{m_b^R} \right) + {\cal O}\left( \frac{1}{(m_b^R)^3},
(\alpha_s^R)^2, \frac{\alpha_s^R}{m_b^R} \right) \right].
\label{eq:Gamma}\end{eqnarray}
Here we have marked all quantities depending on the renormalization
prescription by the superscript ``R''. $f_1$ and $f_2$ are
phase--space factors given by
\begin{eqnarray}
f_1(x) & = & 1-8x^2+8x^6-x^8-24x^4 \ln x,\nonumber\\
f_2(x) & = & 1-\frac{8}{3}\,x^2+8x^4-8x^6+\frac{5}{3}\,x^8 + 8 x^4\ln
x.
\end{eqnarray}
$g^{\text{on--shell}}$ was first analytically calculated
in Ref.\ \cite{nir}:
\begin{eqnarray}
f_1(x)g^{\text{on--shell}}(x) & = & -(1-x^4)\!\left(
\frac{25}{4} - \frac{239}{3}\,x^2+\frac{25}{4}\,x^4\right) + 2 x^2
\left( 20 + 90 x^2 -\frac{4}{3}\,x^4 +
\frac{17}{3}\,x^6\right)\ln x\nonumber\\
& & {}+4 x^4 (36+x^4)\ln x + (1-x^4)\!\left(
\frac{17}{3}-\frac{64}{3}\, x^2 + \frac{17}{3}\,x^4\right)\!
\ln(1-x^2)\nonumber \\
& & {}-8(1+30x^4+x^8)\,\ln x \ln (1-x^2) - (1+16x^4+x^8)\{
6\text{Li}_{2}(x^2) -\pi^2\}\nonumber\\
& & {}-32x^3(1+x^2)\!\left\{ \pi^2 - 4\text{Li}_2 (x) + 4 \text{Li}_2 (-x) - 4
\ln x \ln\, \frac{1-x}{1+x}\right\}.
\end{eqnarray}
$g^{\overline{\text{MS}}}$ can be obtained from (\ref{eq:Gamma}) by
expressing all scheme--dependent quantities in the
$\overline{\text{MS}}$ scheme. To first order in $\alpha_s$, we thus
only need the relation
\begin{equation}
m_b^{\text{pole}} = m_b(\mu)\left\{1+\frac{\alpha_s}{\pi}\left(
\frac{4}{3} + \ln\frac{\mu^2}{m_b^2}\right) + {\cal O}(\alpha_s^2) 
\right\}.
\end{equation}
Insertion in (\ref{eq:Gamma}) yields
\begin{equation}
g^{\overline{\text{MS}}}(x) = g^{\text{on--shell}}(x) + 3 x \ln x 
\,\frac{d\ln f_1(x)}{dx} - 10 + \frac{15}{2}\, \ln \frac{m_b^2}{\mu^2}.
\end{equation}
Numerically, we find the values given in Tab.~\ref{tab:1}.
{}From the comparison of $g^R$ in both schemes, it becomes evident that
higher order corrections in $\alpha_s$ are of paramount importance if
one is willing to obtain reliable predictions. The effect of higher
order terms is conventionally estimated by allowing the
renormalization scale $\mu$ to vary within, say, $m_b/2\leq \mu\leq
2m_b$. Let us compare this with the possible improvement achievable 
by a future calculation of the term in $\alpha_s^2$. It is of the 
generic form
\begin{equation}\label{eq:alphaquadrat}
\frac{\alpha_s^2}{\pi^2}\left( c_1\ln^2\,\frac{m_b^2}{\mu^2} + c_2 \ln
\frac{m_b^2}{\mu^2} + c_3\right),
\end{equation}
where both $c_1$ and $c_2$ are completely determined by the terms of
lower order in $\alpha_s$ and only $c_3$ remains to be calculated.
Expanding the anomalous dimension of the running
$\overline{\text{MS}}$ mass as
\begin{equation}
\gamma^m = \gamma_0^m\,\frac{\alpha_s}{4\pi} + \gamma_1^m
\,\left(\frac{\alpha_s}{4\pi}\right)^2 + {\cal O}(\alpha_s^3)
\end{equation}
and the QCD $\beta$--function as
\begin{equation}
\beta = -g\left\{ \beta_0\,\frac{\alpha_s}{4\pi} +
\beta_1\left(\frac{\alpha_s}{4\pi}\right)^2 + {\cal
O}(\alpha_s^3)\right\},
\end{equation}
we find\footnote{The numerical values of the coefficients $\gamma_i^m$
and $\beta_i$ can be found in any good textbook on QCD.}
\begin{eqnarray}
c_1 & = & \frac{5}{128}\,\gamma_0^m (5\gamma_0^m-2\beta_0),\nonumber\\
c_2 & = & \left. {}-\frac{5}{32}\,\gamma_1^m +
\frac{1}{6}\left(\beta_0+\frac{5}{2}\,\gamma_0^m\right)
g^{\overline{\text{MS}}}(x)\right|_{\mu = m_b}.\label{eq:xxx}
\end{eqnarray}
In Fig.~\ref{fig:scalesGammac} we illustrate the resulting 
scale--dependence of the branching ratio $\tau_D\,\Gamma(D\to X e
\nu)$ for $1\,\text{GeV}\leq \mu \leq 2\,\text{GeV}$. 
We show both the next--to--leading order (NLO) result (\ref{eq:Gamma}) and
the NNLO result with the $\alpha_s^2$ terms (\ref{eq:alphaquadrat})
included where the unknown
constant $c_3$ is varied in the interval $[-10,10]$. As can be read
off the figure, the scale--variation in the NLO result
reproduces approximately the inherent uncertainty from the unknown
constant $c_3$. Still, the remaining scale--uncertainty is noticeable
and limits the accuracy achievable in the determination of $m_c(m_c)$
from $\Gamma(D\to Xe\nu)$. We illustrate that point in
Fig.~\ref{fig:mc} where $m_c(m_c)$ as determined from $\Gamma(D\to
Xe\nu)$ via Eq.\ (\ref{eq:Gamma}) is plotted as a function of the
renormalization scale $\mu$ and for different values of the input
parameters $\lambda_1$ (Fig.~\ref{fig:mc}(a)), $B(D\to Xe\nu)$
(Fig.~\ref{fig:mc}(b)), $m_s(1\,\text{GeV})$ (Fig.~\ref{fig:mc}(c)),
and $\Lambda^{(4)}_{\overline{\text{MS}}}$
(Fig.~\ref{fig:mc}(d)). In Fig.~\ref{fig:mc}(e) we also show
the possible effect of $1/m_Q^3$ corrections estimated by adding to 
and subtracting from (\ref{eq:Gamma}) the $1/m_Q^2$ terms to the power
$3/2$. As input parameters we use\footnote{The experimental errors on 
$\tau_{D^+}$, $|V_{cs}|$, and $|V_{cd}|$ are so small, that the 
combined error of all experimental quantities is completely determined
by the error of the branching ratio, and thus can safely be
neglected.} $B(D\to Xe\nu) = 0.172\pm 0.019$, $\tau_{D^+} = 1.066\,
\text{ps}$, $|V_{cs}| = 0.9743$, $|V_{cd}| = 0.221$ \cite{PD}, 
$m_s(1\,\text{GeV}) = (0.20\pm 0.05)\,\text{GeV}$, cf.\ e.g.\ \cite{PD},
and $\Lambda_{\overline{\text{MS}}}^{(4)} = (300\pm 50)\,\text{MeV}$
\cite{PD}. In contrast to the analysis done in \cite{phenotypen1},
we have varied $\mu$ in the range $1\,\text{GeV}\leq \mu\leq
2\,\text{GeV}$ only, since we believe that the perturbative
expansion becomes highly unreliable at smaller scales. We cannot follow
the arguments of Luke and Savage that $\mu$ could be associated with a
``typical energy'' of the emitted lepton pair and be as small as
$m_c/3 \approx 0.5\,\text{GeV}$! In our approach, $\mu$ has to be
identified with the matching scale of HQET to QCD and cannot be attributed
any physical meaning. The only (loose) condition to be imposed on
$\mu$ is that it should be of order $m_c$. We would also like to mention
that we are rather suspicious about how the scale--uncertainty is handled in
Refs.\ \cite{phenotypen1,phenotypen2,phenotypen3} since in these papers
only the scale in $\alpha_s(\mu)$ is varied (and the
$\overline{\text{MS}}$ scheme used in doing that!), but the quark mass
kept fixed. This corresponds to a scheme where the running of the quark
mass is cut ($\gamma_i^m$ is put zero in (\ref{eq:xxx})), which
is a permissible scheme, but does not seem a very reasonable one. At
least it is {\em not} the correct way how the QCD scale--uncertainty
should be estimated in the on--shell scheme. 

\begin{figure}[h]
\centerline{
\epsfxsize=0.85\textwidth
\epsfbox{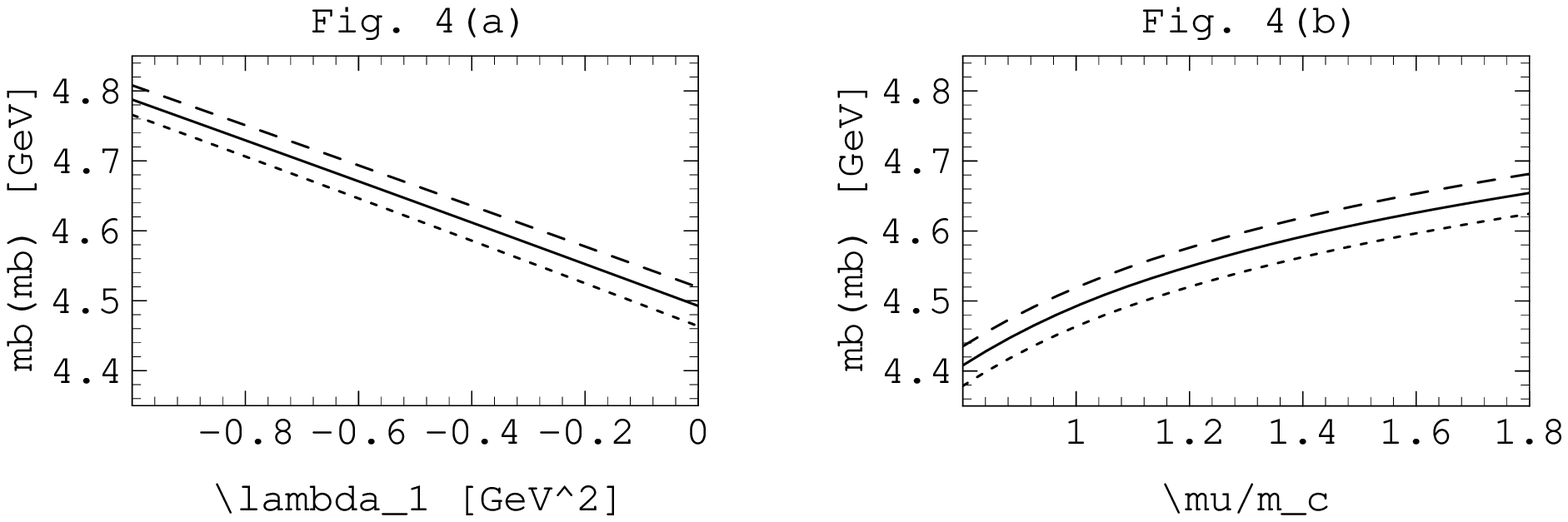}}
\vspace{0.1in}
\caption[]{(a): $m_b(m_b)$ as function of $\lambda_1$ for $\mu = m_c$.
The solid line is calculated with the same set of input parameters as
in Fig.~\protect{\ref{fig:mc}}. The dashed lines correspond to a 
variation of $B(D\to Xe\nu)$ within the experimental error: $B=0.153$ (short
dashes), $B=0.192$ (long dashes). (b): $m_b(m_b)$ as function of
$\mu/m_c$ for $\lambda_1 = 0\,\text{GeV}^2$. The parameter sets used
for the curves are the same as in (a).
}\label{fig:mb}
\vspace{0.2in}
\centerline{
\epsfysize=0.24\textheight
\epsfbox{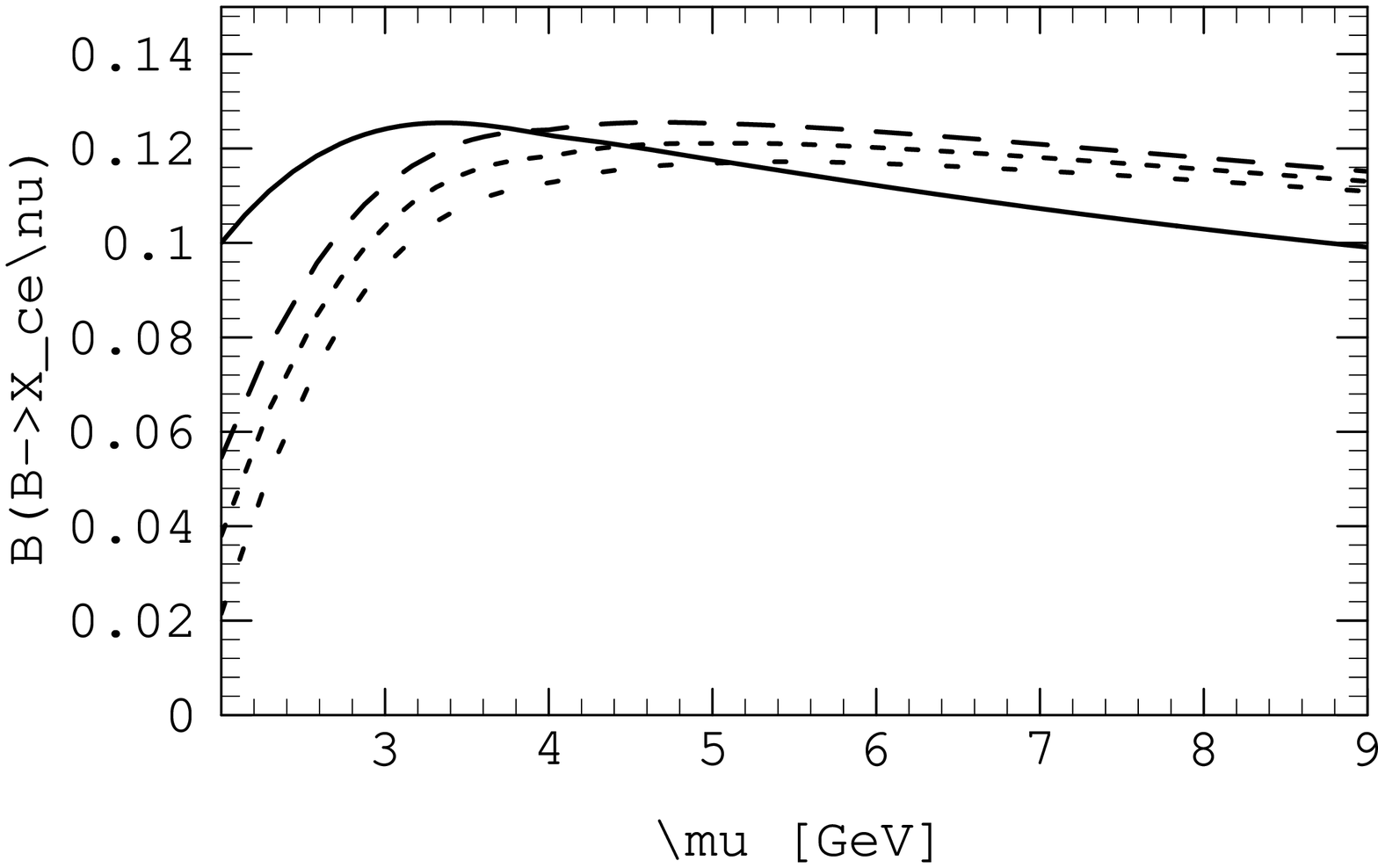}}
\vspace{0.1in}
\caption[]{The branching ratio $B(B\to X_ce\nu)(0.04/V_{cb})^2$ as
function of the renormalization scale $\mu$ for $m_c(m_c)=1.35\, 
\text{GeV}$, $m_b(m_b) = 4.5\,\text{GeV}$, $\lambda_1=0\,\text{GeV}^2$, 
$\lambda_2(m_b) = 0.12\,\text{GeV}^2$. 
Solid line: $B=\tau_B \Gamma$ calculated according to Eq.\
(\protect{\ref{eq:Gamma}}), dashed lines: 
$B$ according to Eq.\ (\protect{\ref{eq:Gamma}}) and with
inclusion of ${\cal O}(\alpha_s^2(\mu))$ corrections according to 
Eq.\ (\protect{\ref{eq:alphaquadrat}}). The different dashed
lines correspond to three different values of $c_3$: $10$, $0$, $-10$ 
(from above).}\label{fig:scalesGammab}
\vspace{0.2in}
\centerline{
\epsfxsize=0.85\textwidth
\epsfbox{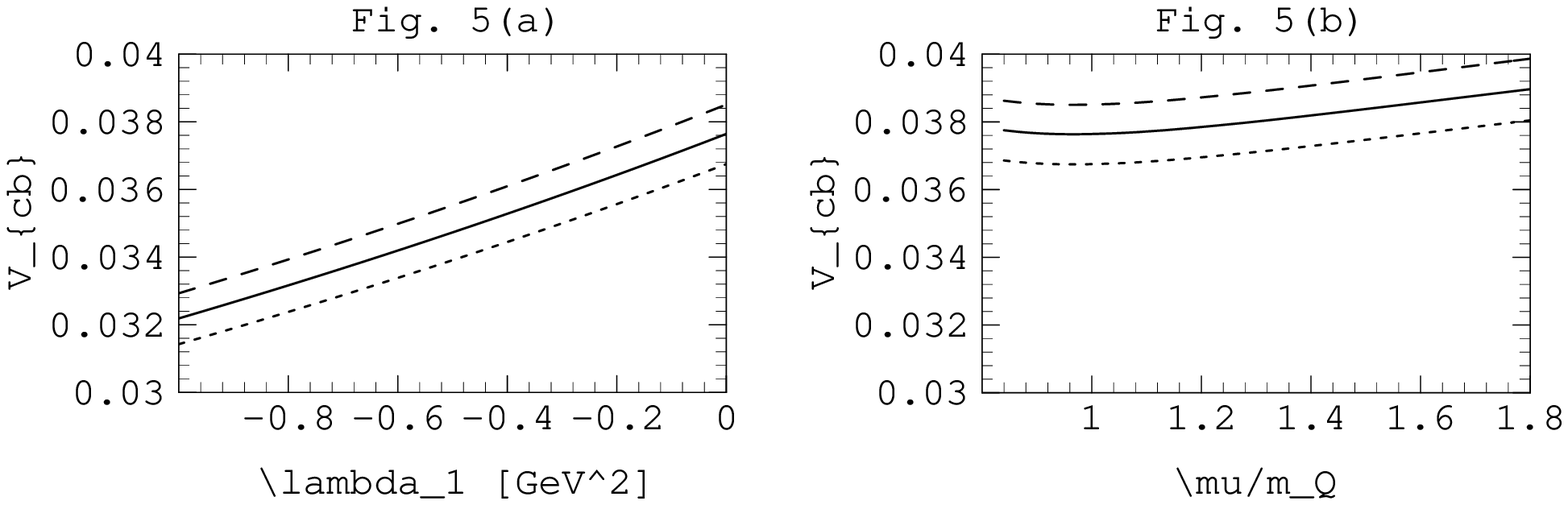}}
\vspace{0.1in}
\caption[]{(a): $|V_{cb}|(\tau_B/1.49\,\text{ps})^{1/2}$ as function  
of $\lambda_1$ for $\mu/m_Q = 1$. The solid line is calculated with 
$B(B\to X_ce\nu)= 0.107$, $B(D\to Xe\nu)= 0.172$, $m_s(1\,\text{GeV}) = 
0.2\,\text{GeV}$ and $\Lambda^{(4)}_{\overline{\text{MS}}}
= 300\,\text{MeV}$. The dashed lines correspond to a variation of $B(B\to
X_ce\nu)$ within the experimental error: $B=0.102$ (short
dashes), $B=0.112$ (long dashes). (b):
$|V_{cb}|(\tau_B/1.49\,\text{ps})^{1/2}$ as function of the ratio 
$\mu/m_Q$ for $\lambda_1 = 0\,\text{GeV}^2$. The parameter sets used
for the curves are the same as in (a).
}\label{fig:Vcb}
\end{figure}

{}From Fig.~\ref{fig:mc} we find  that $m_c(m_c)$ is clearly 
most sensitive to the choice of $\mu$.
In particular, we note that a variation of $\lambda_1$ within the
conservative range $[-0.7,0]\,$GeV$^2$ affects the value of $m_c(m_c)$
only as much as a variation of $B(D\to Xe\nu)$ within one standard
deviation and nearly as much as a variation of the value of the
strange quark mass by $50\,$MeV. These observations seem to cast some
doubts on the procedure employed in 
\cite{phenotypen3} where the determination
of $m_c$ was used to derive bounds on $\lambda_1$ and
$\bar\Lambda$. Taking all together, the heavy quark expansion yields
\begin{equation}
m_c^{\overline{\text{MS}}}(m_c) = (1.35\pm 0.20)\,\text{GeV} \quad \text{for\
}1\,\text{GeV}\leq \mu \leq 2\,\text{GeV.}
\end{equation}
We next use the information gained from the decay of the D meson 
to determine $m_b$. To that end, we use Eq.\ (\ref{eq:massen}) and the
relation 
\begin{equation}
\bar\Lambda_c - \frac{4}{3}\,\frac{\alpha_s(m_c)}{\pi}\,m_c(m_c) =
\bar\Lambda_b - \frac{4}{3}\,\frac{\alpha_s(m_b)}{\pi}\,m_b(m_b)
\end{equation}
which is valid in the $\overline{\text{MS}}$ scheme and can be
obtained from Eq.\ (\ref{laq}). The results are 
displayed in Fig.~\ref{fig:mb}, both for $m_b$ as function of
$\lambda_1$ with fixed scale $\mu=m_c$, Fig.~\ref{fig:mb}(a), and for 
fixed $\lambda_1=0\,\text{GeV}^2$ as function of the scale $\mu/m_c$,
Fig.~\ref{fig:mb}(b). Again
we demand $\mu\geq 1\,\text{GeV}$, which yields $\mu/m_c\gtrsim
0.8$. The dashed lines give the uncertainty of $m_b$ due to the
experimental error in $B(D\to Xe\nu)$. Once more we observe a strong
dependence on the renormalization scale $\mu$, which is however milder
than for $m_c$ due to the higher scales involved. In contrast to the
$c$ quark mass there is also a sizable dependence of $m_b$
on $\lambda_1$ which is of the same order as the $\mu$--dependence and
originates in the relation (\ref{eq:massen}). We obtain
\begin{equation}
m_b(m_b) = (4.6\pm 0.1\pm 0.1\pm 0.1)\,\text{GeV},
\end{equation}
the first error being due to the dependence on $\lambda_1$, the
second one due to $\mu$--dependence, and the third one combines the 
dependence on the $s$ quark mass, the branching ratio $B(D\to Xe\nu )$
and $\Lambda_{\overline{\text{MS}}}^{(4)}$. 
 
In determining $V_{cb}$, we in addition need the experimental
branching ratio for $B\to X_c e \nu$, $B(B\to X_c e \nu) = (10.7\pm
0.5)\% $, quoted in \cite{PD}, and the B lifetime $\tau_B$, where we
use the most recent value $\tau_B = 1.49\,\text{ps}$ quoted in 
\cite{cornell}. In view of the changes the value of
$\tau_B$ has experienced in recent years, cf.\ \cite{PD}, we prefer
not to include its experimental uncertainty in the error analysis, but
to give our results for the above fixed value of $\tau_B$.
In Fig.~\ref{fig:scalesGammab} we show $B(B\to X_c e \nu)$ as function
of $\mu$ with $V_{cb} = 0.04$, $m_c(m_c)=1.35\,\text{GeV}$, $m_b(m_b) =
4.5\,\text{GeV}$, $\lambda_1=0\,\text{GeV}^2$ and $\lambda_2(\mu_b) =
0.12\,\text{GeV}^2$. Analogously to Fig.~\ref{fig:scalesGammac}, the
solid line obtained from Eq.\ (\ref{eq:Gamma}) includes  
${\cal O}(\alpha_s)$ terms with varying $\mu$, whereas the dashed lines 
also include the ${\cal O}(\alpha_s^2)$ corrections
(\ref{eq:alphaquadrat}) with $c_3$ varied within $[-10,10]$. Compared
with Fig.~\ref{fig:scalesGammac}, there is still a
scale--dependence visible, but it is weaker due to the higher scales
involved. Still, the possible effect of a future calculation of $c_3$
can reliably be mimiced by varying $\mu$ within $m_b/2\lesssim \mu
\lesssim 2 m_b$. 

We are now in a position to determine $|V_{cb}|$. Inserting correlated
values of $m_c$ and $m_b$ into Eq.\ (\ref{eq:Gamma}) and comparing
with the experimental branching ratio $B(B\to X_c e \nu)$ we obtain
the values shown in Fig.~\ref{fig:Vcb}. An
analysis of the different sources of uncertainties shows that they are
dominated by the dependence on $\lambda_1$, which proves the
strongest one, and on $B(B\to X_c e \nu)$. In Fig.~\ref{fig:Vcb}(a) we
thus plot $|V_{cb}|$ as function of $\lambda_1$ for fixed scales
$\mu_b/m_b \equiv \mu_c/m_c \equiv 1$ and in Fig.~\ref{fig:Vcb}(b)
as function of the scale $\mu/m_Q$ for fixed $\lambda_1 =
0\,\text{GeV}^2$; we also show the values of $|V_{cb}|$ resulting
from a variation of $B(B\to X_c e \nu)$ within one standard deviation
(dashed lines). For the convenience of the reader, the numerical values
are given explicitly in Tab.~\ref{tab:Vcb}. 
The strong dependence of $|V_{cb}|$ on $\lambda_1$ is
entirely due to the behavior of $m_b$ as function of $\lambda_1$ since
the explicit $\lambda_1/m_b^2$ terms present in Eq.\ (\ref{eq:Gamma})
give only tiny contributions. The rather marginal scale--dependence  
of $|V_{cb}|$ visible
in Fig.~\ref{fig:Vcb}(b) can be explained in the following way. In
Tab.~\ref{tab:scalinv} we give for different scales $\mu/m_Q$ the 
values of the quark masses, their ratio that enters the phase--space
factor, the value of the latter one, the branching ratio divided by
phase--space and the branching ratio itself, calculated with
$|V_{cb}|=0.04$. Although the quark masses differ rather drastically,
the ratio $m_c(\mu_b)/m_b(\mu_b)$ is not very sensitive to $\mu_b$ and
also the phase--space factor $f_1$ varies by only about 10\% in the
range $0.8\leq \mu_b/m_b\leq 1.8$. The branching ratio divided by
phase--space as well as the branching ratio itself are even less
sensitive to the scale. We thus find that the change in $m_b(\mu_b)$ is
nearly completely compensated by a corresponding change in the terms
in $\alpha_s(\mu_b)$ which is far from being trivial. From the figures 
we read off
\begin{equation}\label{eq:resVcb}
|V_{cb}|\left(\frac{\tau_B}{1.49\,\text{ps}}\right)^{1/2} = 0.036\pm
0.002 \pm 0.001 \pm 0.002,
\end{equation}
where the first error includes the dependence on $\lambda_1$ in the
interval $[-0.7,0]\,\text{GeV}^2$, the second one the
scale--uncertainty, and the third one all other uncertainties in the 
parameters, i.e.\ in the branching ratios, $m_s$, and 
$\Lambda_{\overline{\text{MS}}}^{(4)}$. In principle, there is also an
uncertainty due to the scheme--dependence of our analysis which is
however difficult to estimate. If the on--shell scheme was
trustworthy, we could take the difference of our results and those of
\cite{phenotypen1,phenotypen2} as rough estimate for that ``hidden''
theoretical error. For lack of any other calculation in a reasonable
scheme we are however forced to leave that point open to future
investigations. The above value of $|V_{cb}|$ has to be compared with 
\begin{eqnarray}
|V_{cb}|\left(\frac{\tau_B}{1.49\,\text{ps}}\right)^{1/2} &= &
(0.046\pm 0.008)\quad \text{\protect{\cite{phenotypen1}},}\nonumber\\
|V_{cb}|\left(\frac{\tau_B}{1.49\,\text{ps}}\right)^{1/2} & \approx &
0.042\quad\text{\protect{\cite{phenotypen2}},}\nonumber\\
|V_{cb}|\left(\frac{\tau_B}{1.49\,\text{ps}}\right)^{1/2} & = &
0.037\pm 0.007\quad \text{\protect{\cite{cleo}}.}
\end{eqnarray}
The value quoted in \cite{cleo} is the most recent one obtained 
from the spectrum of the {\em exclusive} decay $B\to D^* e\nu$ fitted to 
the Isgur--Wise function using different shapes. We observe that our
result nearly coincides with the one from exclusive decays, but is
smaller than the ones obtained in \cite{phenotypen1,phenotypen2} where
the on--shell scheme was used. Let us close this section with a
short comment on the prospects of a future reduction of the errors in 
(\ref{eq:resVcb}). To begin with, experimentalists are challenged to 
carry out more accurate measurements of the branching ratios,
especially of $B(B\to X_c e\nu)$. Combining with more sophisticated
determinations of $m_s$ \cite{jmprep} and
$\Lambda_{\overline{\text{MS}}}^{(4)}$ will reduce the third error. As
for the second error due to scale--dependence there is no simple remedy
at hand and we hardly can imagine any reliable method to be invented
in the near future to fix the scale except for a calculation of the
${\cal O}(\alpha_s^2)$ corrections to Eq.\ (\ref{eq:Gamma}) which is a
truly formidable task. So the main efforts should be
concentrated on the determination of $\lambda_1$. Although in
\cite{dvf} there was proposed a method of relating $\lambda_1$ to
moments of the so--called ``shape--function'' determining the
end--point region of the lepton-- and photon--spectrum in the decays 
$B\to X_u e\nu$ and $B\to X_s\gamma$, respectively, at present that
approach suffers from missing both measurements and any formulas
going beyond ${\cal O}(\alpha_s^0)$. Yet the introduction of the
shape--function leads to the constraint $\lambda\leq 0\,\text{GeV}^2$
which is in agreement with the results obtained in \cite{BBlast} and 
\cite{gurbound}, but in contrast to the earlier determination 
\cite{nn}, where a positive value
was obtained. The status of the careful analysis done in \cite{BBlast}
with QCD sum rules is unfortunately not beyond any doubt, cf.\
\cite{vir}, so a clarification of that point would be welcome. If the
result of \cite{BBlast} could be confirmed, (\ref{eq:resVcb}) would
read $|V_{cb}|(\tau_B/1.49\,\text{ps})^{1/2} = 0.034\pm 0.001 \pm 0.002$.

\section{Summary and Conclusions}

In applications of the heavy quark expansion of QCD 
to physical quantities involving positive powers of the 
heavy quark mass the proper definition of the mass parameter 
is of supreme phenomenological importance. We have made clear
that the heavy quark mass parameter stemming from the Wilson coefficient in 
the matching of perturbative QCD and HQET Green functions, in a rigorous QCD 
context enters the calculation as the running current mass 
rather than the pole mass. It is evaluated at the  scale at which we perform 
the matching. This scale must be of order of 
the heavy quark mass, but can be varied to estimate the error 
caused by the truncation of the perturbative series. 
The choice of the on--shell scheme, which is widely used also for 
the mass parameters in the Wilson coefficient, corresponds 
to a special choice for $\mu$, which is 
unnaturally low in the case of the $c$ quark.
It is well--known that the heavy quark expansion reproduces in
leading order the parton model. With the proper identification of 
the mass parameter stemming from the Wilson coefficient we also 
recover the familiar renormalization scheme-- and scale--dependence of 
the parton model.

The natural expansion parameter of the HQET Green functions, however, 
is the pole mass, which was recently found to be ambiguous when 
defined beyond finite orders in perturbation theory. 
Since HQET attempts to include non--perturbative effects, a
phenomenologist using HQET has to worry whether this ambiguity may 
be relevant for his physical predictions. We have argued that a
calculation which includes only radiative corrections to a 
finite order to a given term in the $1/m_Q$--expansion, requires
only the calculation of the perturbative pole mass to the same order
in $\alpha_s$. This fact is plausible, but not completely obvious,
because the usual derivation of HQET requires the 
all--orders pole mass in the first step to define the heavy quark
field $h_v$. 

The proper identification of the  mass parameters  becomes
probably most pronounced in the case of semileptonic inclusive decays.
Their decay rate depends on the quark mass in its fifth power and is most
sensitive to any change in the definition of that parameter. 
Here  it likewise becomes rather obvious that the on--shell scheme is,
even apart from the renormalon problem, a rather unnatural one, since
it is always the {\em current mass} that appears in actual
calculations, whereas the pole mass has to be put in afterwards by
hand, adding higher order terms in $\alpha_s$ in an uncontrollable
manner. We have carefully studied the effect of changing the scheme
and scale  
and found that neither the $c$ nor the $b$ quark mass can 
be determined reliably from inclusive decays. In extracting
$|V_{cb}|$, however, the scale--dependence mostly cancels and we found
a reduction of the value of $|V_{cb}|$ by
more than 10\% compared with results obtained in the on--shell scheme
\cite{phenotypen1,phenotypen2}. By a more consistent error analysis,
in particular of the scale--dependence, we also could reduce the
absolute error of $|V_{cb}|$. We also clarified the meaning of the
renormalization scale $\mu$ entering the parton model and its
non--perturbative corrections and identified it with the scale at
which HQET is matched to QCD, a point that was not paid proper
attention to in the previous analyses 
\cite{phenotypen1,phenotypen2,phenotypen3}.

Finally, we suppose that our considerations could also give some hint 
at the solution of the problem of the ``baffling semileptonic
branching ratio of $B$ mesons'' raised in \cite{baffl} and that it
could be worthwile to attack that problem starting from a different 
choice of the renormalization scheme.

\acknowledgments

We thank A.J.\ Buras and E. Bagan for stimulating discussions. 
U.N.\ appreciates clarifying conversations with M. Misiak, 
W. Kilian and T. Ohl. P.B.\ would like to thank M. Shifman for sending her a
preliminary version of the first paper in Ref.\ \cite{privet} prior to
publication. 
We further thank G. Buchalla for carefully reading the manuscript
and for critical remarks.
We are especially grateful to M. Beneke for his thorough analysis of 
an earlier version of this paper and the comprehensive explanation
of the physical consequences of the renormalon problem in the 
definition of the pole mass.

\newpage

\clearpage
\makebox[1cm]{}
\vskip-0.6in

\newpage
\begin{table}
\squeezetable
\begin{tabular}{lddddddddddd}
$x$ & 0.0 & 0.1 & 0.2 & 0.3 & 0.4 & 0.5 & 0.6 & 0.7 & 0.8 & 0.9 & 1.0\\
\tableline
$g^{\text{on--shell}}$ & 3.62 & 3.25 & 2.84 & 2.51 & 2.23 & 2.01
& 1.83 & 1.70 & 1.59 & 1.53 & 1.50\\
$g^{\overline{\text{MS}}}$ & $-$6.38 & $-$7.79 & $-$9.94 & $-$12.1 & 
$-$14.2 & $-$16.0 & $-$17.8 & $-$19.4 & $-$20.9 & $-$22.2 & $-$23.5\\
\end{tabular}
\caption[]{Radiative corrections to the semi--leptonic free quark
decay, Eq.~(\protect{\ref{eq:Gamma}}), in the on--shell and in the
$\overline{\text{MS}}$ scheme.}\label{tab:1}
\end{table}

\begin{table}
\begin{tabular}{ldddddddd}
$\lambda_1\,[\text{GeV}^2]$ & $-$0.7 & $-$0.6 & $-$0.5 & $-$0.4 & 
$-$0.3 & $-$0.2 & $-$0.1 & 0\phantom{.0000}\\ \tableline
$B=0.102$ & 0.0329 & 0.0334 & 0.0339 & 0.0344 & 0.0350 & 0.0356 &
0.0362 & 0.0367 \\
$B=0.107$ & 0.0337 & 0.0342 & 0.0347 & 0.0353 & 0.0358 & 0.0364 & 
0.0370 & 0.0376 \\
$B=0.112$ & 0.0344 & 0.0350 & 0.0355 & 0.0361 & 0.0367 & 0.0373 & 
0.0379 & 0.0385 \\
\end{tabular}
\caption[]{$|V_{cb}|(\tau_B/1.49\,\text{ps})^{1/2}$ for several values
of $\lambda_1$ and the branching ratio $B(B\to X_ce\nu)$. The same input
parameters are used as in Fig.~\protect{\ref{fig:Vcb}(a)}.}\label{tab:Vcb}
\end{table}

\begin{table}
\begin{tabular}{ldddddd}
$\mu/m_Q$ & $m_c(m_c)\,[\text{GeV}]$ & $m_b(m_b)\,[\text{GeV}]$ & 
$m_c(\mu_b)/m_b(\mu_b)$ & $f_1$ & $\Gamma \tau_B/f_1$ &
$\Gamma \tau_B$ \\  \tableline
0.8 & 1.28 & 4.41 & 0.21 & 0.72 & 0.171 & 0.123 \\
1.3 & 1.43 & 4.57 & 0.23 & 0.67 & 0.182 & 0.122 \\
1.8 & 1.51 & 4.65 & 0.24 & 0.64 & 0.179 & 0.116 \\
\end{tabular}
\caption[]{Quark masses, their ratio, phase space $f_1$ and branching
ratio $\Gamma \tau_B$ for different values of the scale $\mu/m_Q$. In
the last two columns we have put $|V_{cb}|=0.04$.}\label{tab:scalinv}
\end{table} 

\end{document}